\documentclass[prl,reprint,nobibnotes]{revtex4-1}

\usepackage{color}

\usepackage{graphicx}
\usepackage{booktabs}
\usepackage{bm}
\usepackage{longtable}
\usepackage{amsmath}
\usepackage{dcolumn}
\usepackage{placeins}
\usepackage{expdlist}

\usepackage{siunitx}

\begin{document}

\bibliographystyle{apsrev}

\title {Identification of object composition with Magnetic Inductive Tomography}

\author{R. Gartman }
\affiliation{National Physical Laboratory, Hampton Road, Teddington TW11 0LW, United Kingdom}
\author{W. Chalupczak}
\email{The author to whom correspondence may be addressed: witold.chalupczak@npl.co.uk}
\affiliation{National Physical Laboratory, Hampton Road, Teddington TW11 0LW, United Kingdom}

\date{\today}

\begin{abstract}
The inductive response of an object to an oscillating magnetic field reveals information about its electrical conductivity and magnetic permeability.  Here we introduce a technique that uses measurements of the angular, frequency and spatial dependence of the inductive signal to determine object composition.  Identification is performed by referencing an object's inductive response to that of 
materials with mutually exclusive properties such as copper (high electric conductivity, negligible magnetic permeability) and ferrite (negligible electric conductivity, high magnetic permeability). The technique uses a sensor with anisotropic sensitivity to discriminate between the different characters of the eddy current and magnetisation driven object responses. Experimental validation of the method is performed through Magnetic Induction Tomography measurement with a radio-frequency atomic magnetometer. Possible applications of the technique in security screening devices are discussed.
\end{abstract}

\maketitle

\section{Introduction} 
\begin{figure}[h!]
\includegraphics[width=\columnwidth]{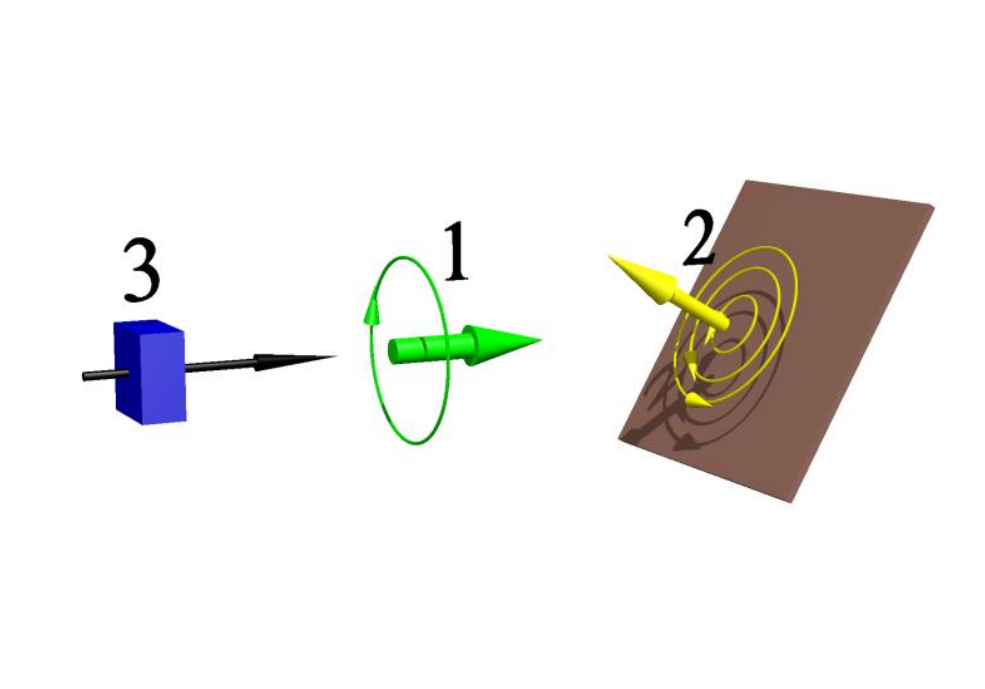}
\caption{Generic components of a Magnetic Induction Tomography measurement setup. The primary rf field (green arrow) produced by the rf coil (1) causes the inductive response of the object (2) in the form of a secondary field (yellow arrow). The signal is recorded by a sensor (3). Here, the radio-frequency atomic magnetometer as a sensor monitors only the components of the secondary field orthogonal to the sensor axis (black arrow).}\label{fig:Setup}
\end{figure}

Magnetic Induction Tomography (MIT) measurements rely on the inductive coupling between a radio-frequency (rf) magnetic field, the so-called primary rf field, and the object of interests, Fig. ~\ref{fig:Setup}, \cite{Griffiths2001, Ma2017}. As a result of the coupling an object response is produced in the form of a secondary rf field. 
For objects whose response is dominated by electrical conductivity, eddy currents induced by the primary rf field produce the secondary rf field that opposes the driving field. This leads to dissipation of the primary rf field and reduced field penetration within the object. 
When the response is dominated by magnetic permeability the primary field creates within the object a magnetisation oscillating in phase with the driving field. 
In general, any object shows some level of electric conductivity and magnetic permeability. The secondary rf field reflects the character of the dominating property but the measured amplitude and phase of the inductive response depends on relative ratio between the electric conductivity and magnetic permeability, which in principle indicates the composition of the object.

MIT provides a portfolio of measurements addressing a wide range of contemporary challenges in applied physics. In the area of non-destructive testing (NDT), inductive measurements enable detection of defects either covered by insulation or concealed within the object structure \cite{Auld1999, Perez2004, Deans2017, Yoshimura2019, Bevington2020a}. Immediate applications of the technology lie in the energy sector where corrosion under insulation is responsible for a significant fraction of the losses in the transport and storage of oil and gas. Implementations of MIT in object detection and surveillance include imaging through barriers and in turbulent underwater environments that prevent the use of visual or ultrasound technology.

The use of an rf atomic magnetometer as the sensor in MIT brings superior sensitivity \cite{Savukov2005, Chalupczak2012, Deans2016, Deans2018a} as well as a range of functionalities such as the ability to obtain vector measurements \cite{Bevington2019, Bevington2019b}, high bandwidth in self-oscillating mode \cite{Bevington2019c, Bevington2020}, and tunability over a wide frequency range without compromise of performance \cite{Wickenbrock2016, Bevington2020c}. Measurement of the object response with an rf atomic magnetometer relies on monitoring the change in the amplitude and phase of the rf resonance recorded with the magnetometer while scanning across the material, Fig.~\ref{fig:Setup}. 

Our studies so far have been focused on the implementation of rf atomic magnetometers for inductive tomographic mapping in scenarios of defect detection and object surveillance. These include the demonstration of tomographic mapping of material thinning in steelwork, which represents the detection of corrosion under insulation \cite{Bevington2018}. The imaging provides us with a vector measurement (2D) of the secondary field, while the orientation of the sensor axis, i.e. direction of the bias magnetic field, defines which field components are being monitored \cite{Bevington2019, Bevington2019b}. With the sensor and the primary rf field coil axes parallel to each other the magnetometer signal represents only the amplitude of the secondary field which leads to increased image contrast. Implementation of a spin maser, an rf atomic magnetometer operating in self-oscillating mode, increases the data acquisition rate and makes the sensor immune to variations in the ambient magnetic field \cite{Bevington2019c, Bevington2020}. This increase in measurement bandwidth comes at the price of the reduction of the part of the defect/ object signature that results the phase matching condition in the sensor feedback loop. The application of a pair of the primary rf field coils with opposite polarity, a dual frequency spin maser \cite{Bevington2020b} or an external phase scan can solve this issue.
Whilst this inductive tomographic mapping can provide information about the depth and spatial extent of a defect or object, it requires a scan of the sensor over the area of interest. Although the scan time can be optimised there is a category of scenarios, such as security screening, that requires rapid measurements that are possible at a single location and determine whether more detailed screening is required. Usually, this decision is based on the ability to discriminate between different types of materials.

In this paper, we present a technique that can potentially help determine object composition and hence reduce measurement duration. It combines measurements of the angular, frequency and spatial dependence of the signal with comparisons of the object's inductive response to those of reference materials with mutually exclusive properties such as copper (high electric conductivity, negligible magnetic permeability) and ferrite (negligible electric conductivity, high magnetic permeability). While the rf signal frequency dependence has been shown to provide discrimination between different objects, the demonstration was limited to a narrow class of purely conductive materials and required a series of extra measurements for calibration \cite{Wickenbrock2016, Deans2018c}. The discrimination discussed in \cite{Wickenbrock2016, Deans2018c} was based on the direct dependence of the inductive signal on the electrical conductivity of objects with negligible magnetic permeability. Identification of objects whose inductive response, the secondary rf field, results from both eddy currents and magnetisation components is more complex. Moreover, as we demonstrate, the signal depends on the experiment geometry and the object shape, complicating both the measurements and data analysis. We present frequency and angular dependence measurements that are performed at single spot above the object, which can reduce the screening time. The focus of this paper is on validation of the technique, i.e. demonstration of a series of measurements that can provide discrimination between objects made of different materials. Analysis of the data for practical application can be improved by the introduction of various metrics, such as those based on machine learning. We demonstrate two methods of inductive image analysis that can assist in identification of object composition, the first based on the signal amplitude's frequency dependence and the second using the amplitude integrated over the entire image area.

\section{Experimental setup} 
The measurements described here are performed with a radio-frequency atomic magnetometer operating in a magnetically unshielded environment \cite{Bevington2018, Bevington2019, Bevington2019b}. For the purposes of the techniques described here, the technical details of the atomic magnetometer are not essential. A description of the sensor and instrumentation is presented elsewhere \cite{Bevington2019, Bevington2019b} and we limit discussion of the sensor to the enumeration of its major components. Our rf atomic magnetometer instrumentation includes three major subsystems: lasers, caesium atomic vapour contained in a paraffin-coated cell and the detection. The cell is kept at ambient temperature (atomic density $n_{\text{Cs}}=0.33 \times10^{11} \text{cm}^{-3}$) in a static magnetic bias field, created by a set of nested, orthogonal, square Helmholtz coils. The strength of the bias field defines the operating (Larmor) frequency of the sensor. The laser system produces two beams. A circularly polarised pump beam, frequency stabilized to the $6\,^2$S$_{1/2}$ F=3$\rightarrow{}6\,^2$P$_{3/2}$ F'=2 transition (D2 line, $\SI{852}{\nano\meter}$) propagates along the direction of the bias magnetic field. It creates a population imbalance within the ensemble of caesium atoms. A probe laser, whose frequency is tuned $\SI{2.75}{\giga\hertz}$ below the $6\,^2$S$_{1/2}$ F=3$\rightarrow{}6\,^2$P$_{3/2}$ F'=2 transition, propagates in a direction orthogonal to the pump beam. It monitors the atomic signal created by the coupling of the atoms and the rf magnetic fields (i.e. atomic coherence). The primary rf field, oscillating at the sensor operating frequency, is produced by a coil located in the vicinity of the object. Lift off, the distance between the primary rf field coil and the object, is between $\SI{2}{\milli\meter}$ and $\SI{20}{\milli\meter}$. The axis of the primary rf field is parallel to the bias field direction. It is important to stress that the atomic magnetometer can sense only the rf magnetic field that is perpendicular to the direction of the bias magnetic field. In the following we refer to the bias field direction as the axis of the sensor. The parallel orientation of the sensor axis and the primary rf field makes the sensor insensitive to the primary rf field \cite{Bevington2019b}. Consequently, the sensor readout, either measured by a lock-in amplifier or recorded by a 2 MS/s data acquisition board, monitors directly the secondary rf field. This simplifies the analysis of the data and makes the normalisation procedure, essential in \cite{Wickenbrock2016}, obsolete.

\begin{figure}[tbp]
\includegraphics[width=\columnwidth]{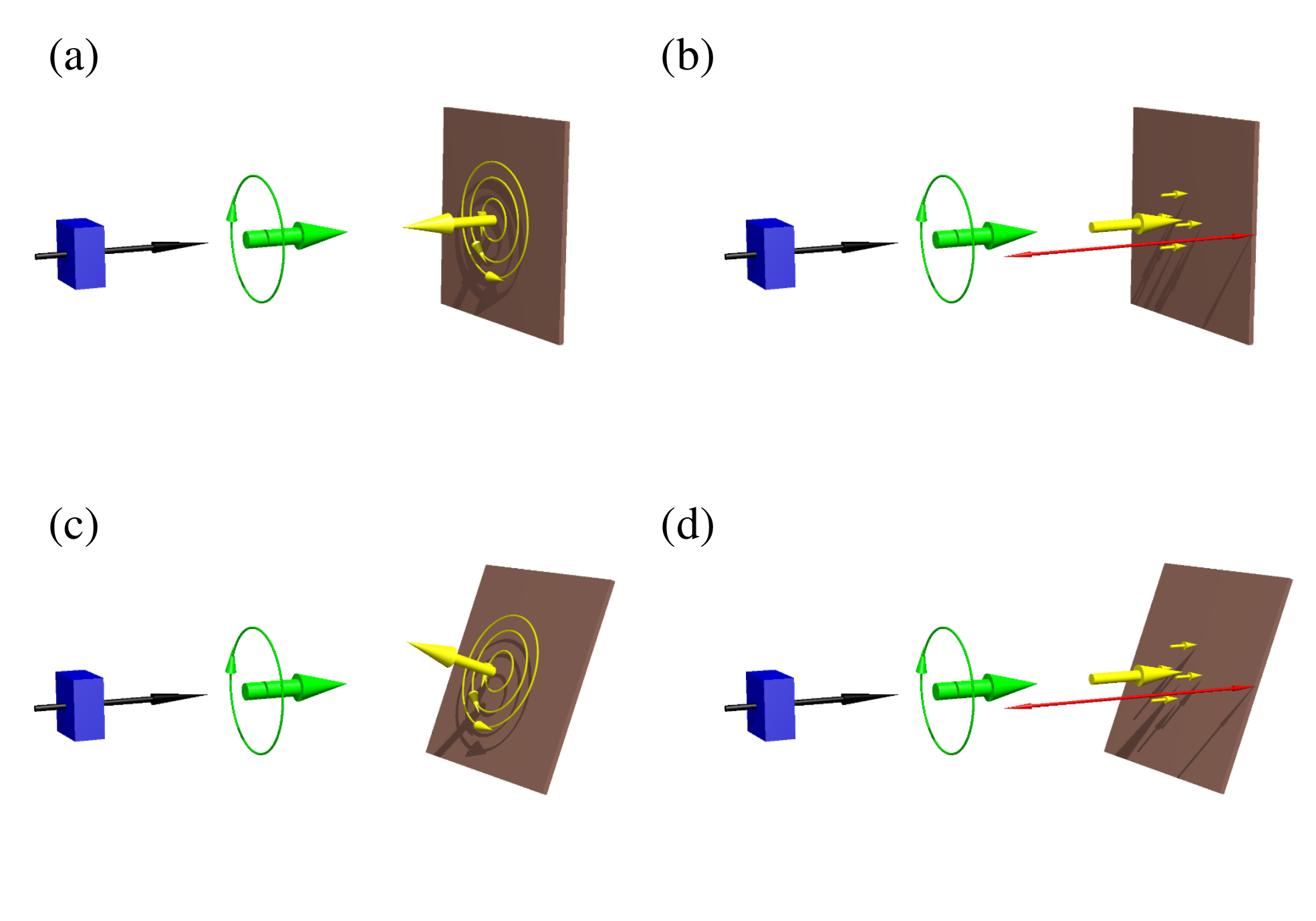}
\caption{The secondary rf field produced by eddy currents (a), (c) and magnetisation (b), (d) in measurement geometries where the normal to the object surface is either parallel (a), (b) or tilted at an angle (c), (d) to the primary rf field direction.  The red arrows in (b) and (d) indicate how we define the lift off in each geometry.}\label{fig:Secondary_field}
\end{figure}

\section{Measurement geometry}
In general, eddy currents and magnetisation induced within the object produce two secondary rf field components that have different amplitude and phase characteristics. Depending on the relative electric conductivity and magnetic permeability values, one of these components dominates the object's inductive response. In this section we show that the measurement geometry, specifically the relative orientation between the normal to the object surface and the sensor axis, can suppress or enhance the contribution to the measurement signal from each component of the secondary field.  This provides a mechanism to distinguish between them.

Due to eddy current driven dissipation, penetration of the rf field within an electrically conductive object can be limited to a thin layer in the immediate vicinity of the surface. In the particular case of an object made from aluminium, the skin depth is $\SI{0.8}{\milli\meter}$ for a primary rf field frequency of $\SI{10}{\kilo\hertz}$. In general it can be expected that for any object the component of the secondary field due to eddy currents is produced in the immediate vicinity of the surface and its direction is parallel to the normal to the surface. This means that the secondary field direction reflects the orientation of the surface and any change in the orientation of the object surface results in a change of the direction of the secondary field. This will manifest itself as a change in the detected rf signal amplitude and phase. In contrast, in objects with negligible electrical conductivity (and hence low rf field dissipation) and high magnetic permeability, such as ferrites, the direction of the secondary field is defined by magnetisation throughout the object. It is parallel to the primary rf field direction regardless of the orientation of the object. Hence, it can be expected that the component of the secondary field produced by magnetisation in any object mirrors the primary rf field direction.

To gain further insight we consider two measurement geometries, the first where the normal to an object's surface is parallel to the primary rf field, Fig.~\ref{fig:Secondary_field} (a)-(b), and the second where there is a non-zero tilt between the two, Fig.~\ref{fig:Secondary_field} (c)-(d). As discussed earlier, the primary rf field direction (green arrow) is parallel to the axis of the sensor (black arrow). The atomic magnetometer used as the sensor is insensitive to rf field components directed along this axis, so the primary rf field does not contribute to the measured signal.
In the first configuration components of the secondary field produced by both eddy currents and magnetisation are parallel to the axis of the sensor, and are in consequence invisible to it. With a non-zero tilt between the axes the direction of the component produced by eddy currents is no longer parallel to the sensor axis, making it visible to the sensor, Fig.~\ref{fig:Secondary_field} (c). The direction of the component produced by magnetisation remains parallel to the sensor axis, and so does not contribute to the detected signal, Fig.~\ref{fig:Secondary_field} (d). In general, with increasing angle between the sensor axis and the normal to the object surface the visibility of the eddy current driven component increases, while that of the magnetisation component doesn't change.

\begin{figure}[tbp]
\includegraphics[width=\columnwidth]{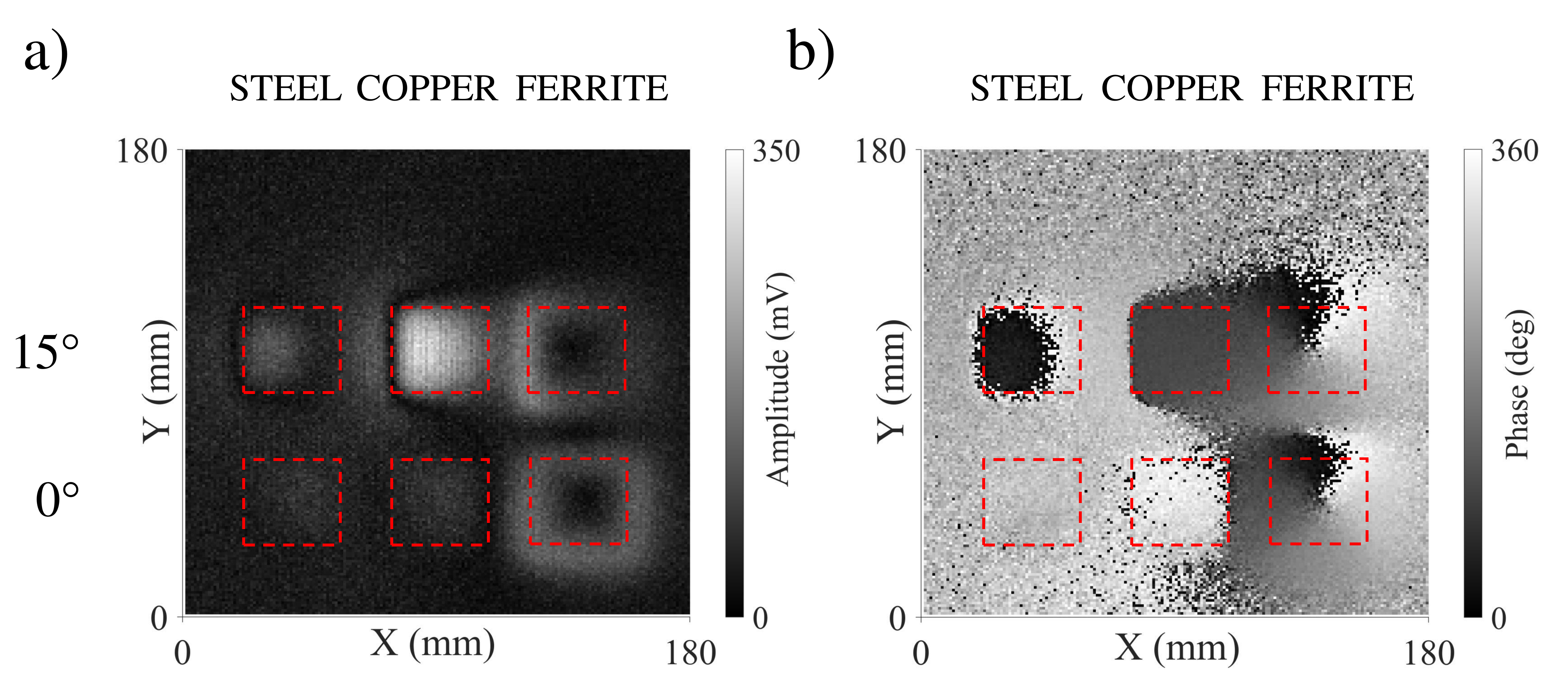}
\caption{Plots of the amplitude (a) and phase (b) of measured rf signals generated by a single scan over pairs of the stainless steel, copper and ferrite $35\times\SI{35}{\milli\meter\squared}$ plates. Red dashed lines mark the positions of the plates. For amplitude and phase images in the lower row the normal to the surface of the plates is parallel to the primary rf field, whilst there is a $15^{\circ}$ tilt between them for images in the upper row. The axis of rotation is directed along the Y axis through the centre of the plates. The image was recorded at an operating frequency $\SI{29}{\kilo\hertz}$. }\label{fig:Tilt_image}
\end{figure}

To illustrate the differences between object responses produced by eddy currents and magnetisation in different measurements geometries two sets of amplitude and phase inductive images were recorded. The images were generated by a single scan over three pairs of stainless steel, copper and ferrite $35\times\SI{35}{\milli\meter\squared}$ plates, marked with red dashed lines in Fig. ~\ref{fig:Tilt_image}. All the plates used in the experiment were $\SI{0.5}{\milli\meter}$ thick, except the ferrite, which was $\SI{2}{\milli\meter}$ thick. The image in Fig. ~\ref{fig:Tilt_image} was recorded with the normal to the object surface either parallel to the primary rf field (lower row) or with a $15^{\circ}$ tilt with respect to it (upper row). 
The lift off, defined as shown in Fig.~\ref{fig:Secondary_field} (b) as the distance from the primary rf coil to the axis of plate rotation, was $\SI{10}{\milli\meter}$. The scans were performed at an operating frequency of $\SI{29}{\kilo\hertz}$. The choice of this particular frequency will be explained in the following section. The images of the ferrite plate represent the case when the inductive response is solely produced by the magnetisation of the object. Both ferrite amplitude images show a dark area produced by the centre of the plate surrounded by a bright square created by the edges. This results from the secondary field component parallel to the plate surface created by the plate edges \cite{Bevington2020d}. Both ferrite phase images show the presence of a vortex, another signature of the plate edges \cite{Bevington2020d}. For this material the amplitude and phase images recorded in different measurement geometries have the same structure and the signals have similar dynamic range, which supports the expectation that the magnetisation orientation is the same regardless of the measurement configuration. The smaller amplitude on right hand side of the image recorded with $15^{\circ}$ tilt between the axes results from the bigger lift off. 
The images of the copper plate represent the case when the inductive response is produced by eddy currents within the object. Images recorded in different geometries differ not only in amplitude but also in phase. The latter confirms that the direction of the secondary field produced by eddy currents depends on the orientation of the object's surface. It is worth pointing out the reversed character of the copper amplitude image recorded at $15^{\circ}$ with respect to ferrite one. The inner part representing the secondary field created by the centre of the plate is bright and is surrounded by a dark square produced by the edges.
The stainless steel represents an object that exhibits both electrical conductivity and a some magnetic permeability. Because the permeability of stainless steel is smaller than that of the ferrite, the signature of the plate edges is small and neither the bright square nor the phase vortex is visible when the plate surface is not tilted.  
With a $15^{\circ}$ tilt between the sensor axis and the normal to the plate surface the amplitude and phase signatures become visible. The similarity of these signatures to those produced by the copper plate indicate that in this case the secondary field also originates from eddy currents.

\begin{figure}[tbp]
\includegraphics[width=\columnwidth]{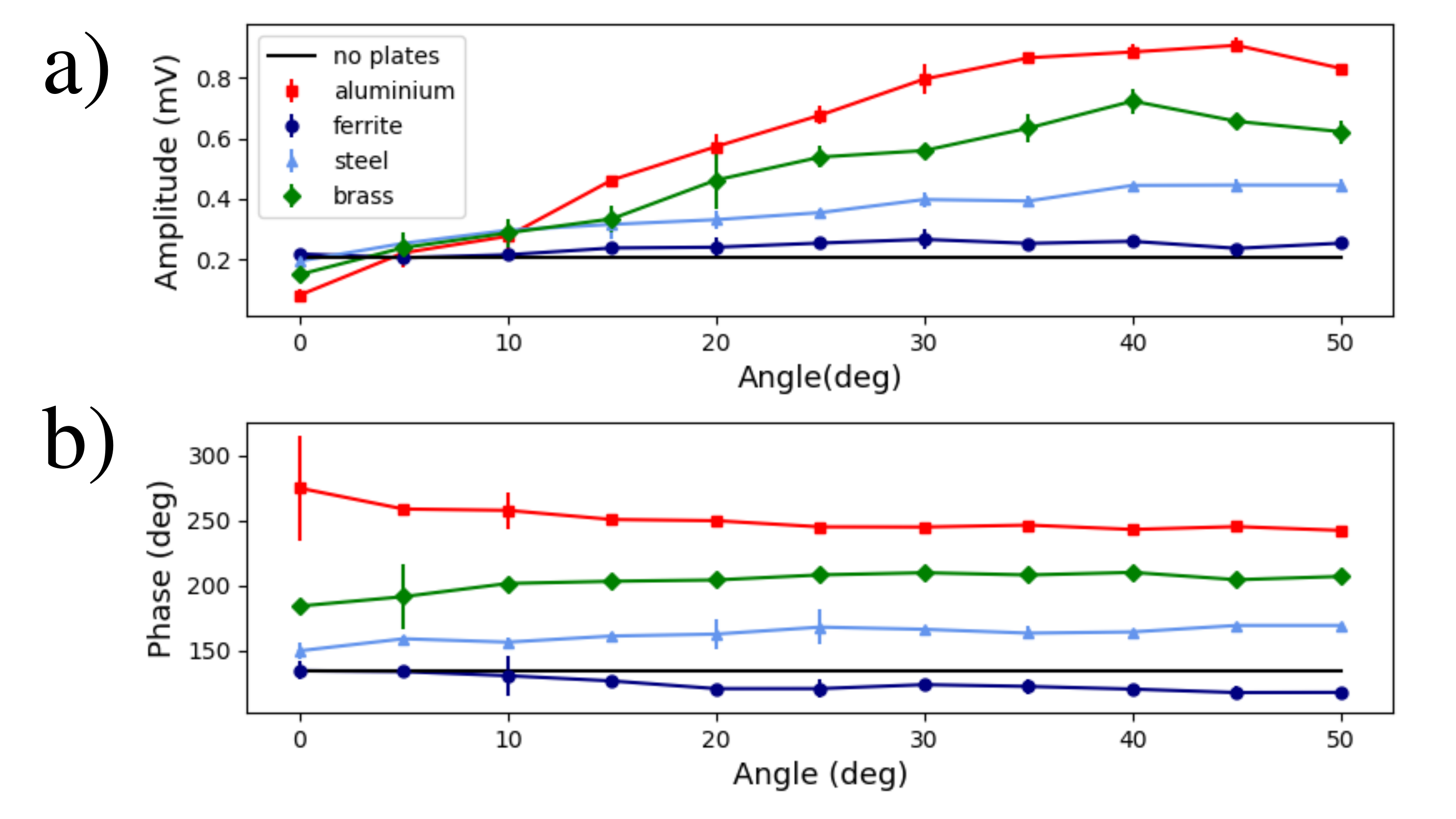}
\caption{Amplitude (a) and phase (b) of the rf signal  as a function of the angle between the normal to the object surface and the sensor axis recorded for the ferrite (dark blue points), stainless steel (light blue triangles), brass (green diamonds) and aluminium (red squares) plates. Lines conecting points serve only as eye guides. For reference the signal level recorded in the absence of the sample is shown with the solid black line. The measurement was performed at an operating frequency $\SI{6.7}{\kilo\hertz}$. The primary rf field coil is located above the centre of the plate.}\label{fig:Tilt}
\end{figure}

Figure ~\ref{fig:Tilt} shows the rf signal amplitude (a) and phase (b) as a function of the angle between the normal to the object surface and the sensor axis recorded at a single location above ferrite (dark blue points), stainless steel (light blue triangles), brass (green diamonds) and aluminium (red squares) plates. The measurement was done by placing the object on a support plate attached to a rotation mount. Care was taken to ensure that the primary rf field coil was located above the axis of rotation, as in Fig.~\ref{fig:Secondary_field}, such that the change of object orientation does not affect its distance to the primary rf field coil.
The amplitude and phase of the signal produced by the ferrite plate does not change significantly with plate orientation, confirming that the secondary field generated by magnetisation mirrors the primary rf field direction. The high magnetic permeability and low electrical conductivity result in signals that are similar to those obtained in the absence of a sample. 
In the case of the aluminium plate, the amplitude of the signal increases in angle range $0^{\circ}-45^{\circ}$. 
The non-zero signal amplitude in the absence of the object results from residual misalignment between the primary rf field and sensor axes.
It is worth reiterating that this is a result of the measurement configuration and magnetisation behaviour, Fig. ~\ref{fig:Secondary_field} (b) and (d), where the sensor axis, indicated by the black arrow, is parallel to the direction of the primary rf field (green arrow) and the secondary rf field (yellow arrow).

The secondary field component due to eddy currents changes direction with plate rotation.  The detected signal is sensitive only to the projection of the secondary field onto the plane perpendicular to the sensor axis, with the amplitude given by the radius of this projected vector and the phase by the radial angle. Here we rotate the plate about an axis that is perpendicular to the sensor axis (Fig.~\ref{fig:Secondary_field}), which changes the radius of the projected vector but not the radial angle.  As a result we see a change in the amplitude of the signal, but no change in phase. 
The difference between the phases measured for ferrite and aluminium plates at any given tilt is about $120^{\circ}$ and reflects the different character of the effect that generates the secondary field. The stainless steel plate possesses significant electrical conductivity and residual magnetic permeability and while the latter dominates for low angles the former becomes visible with increasing angle. This is reflected in the increase of signal amplitude. The lower conductivity of the stainless steel plate is reflected in the lower signal amplitude and phase relative to the aluminium plate when observed at larger tilt angles. The values of the amplitude and phase produced by the brass plate lie between those of stainless steel and aluminium, which is consistent with its intermediate conductivity.
An angular dependence of the amplitude and phase of the signal similar to that presented in Fig.~\ref{fig:Tilt} is observed for operating frequencies above $\SI{4}{\kilo\hertz}$, which confirms that the effect requires eddy current generation limited to the immediate vicinity of the surface.

Because of the measurement configuration, where the sensor has an insensitive axis that is aligned with the primary rf field, the angular dependence of the measured signal amplitude and phase is affected by an object's geometry. In the particular case of a plate the amplitude reaches a minimum for $0^{\circ}$ and $90^{\circ}$ when a surface of the plate faces the rf primary field, because the surface orthogonal to the primary rf field does not contribute to the signal. This can be seen in Fig. ~\ref{fig:Tilt} (a), where the signal amplitude reaches a maximum at $45^{\circ}$ and shows signs of decreasing for bigger angles. It is worth noticing that the thickness of the plate is not important. The same angular dependence of the signal amplitude would be observed in the case of a cubic box. 
One might expect that, for regular shapes the number of minima in the angular dependence of the signal amplitude and the angles at which they occur can provide information about the symmetry of an object.  In the more general case, a proper understanding of how the output of the local measurement depends on object geometry is important in the reconstruction of object shape and composition.

\section{Frequency scan}
\begin{figure}[h!]
\includegraphics[width=\columnwidth]{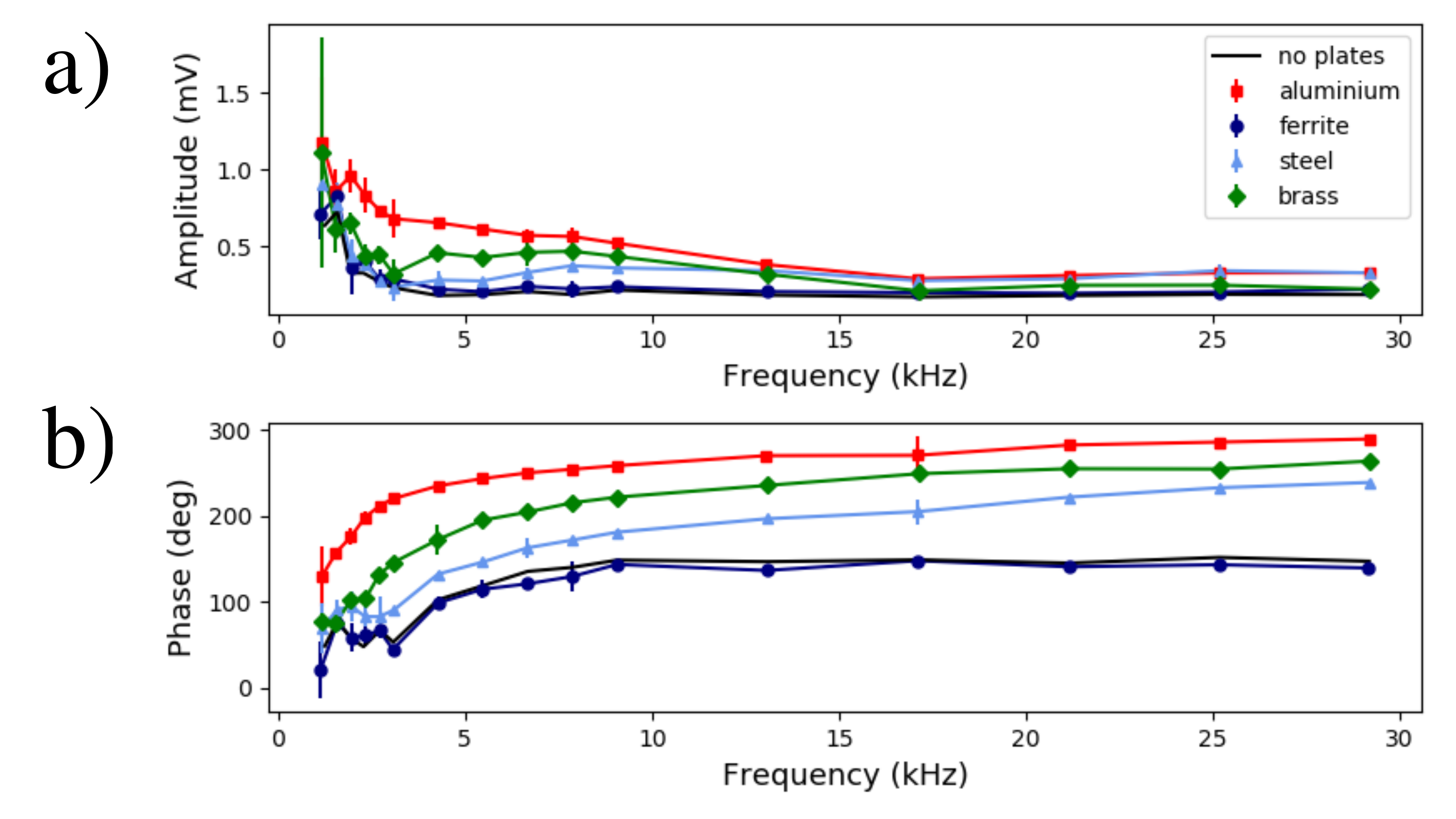}
\caption{Amplitude (a) and phase (b) of the rf signal as a function of the operating frequency recorded for the ferrite (dark blue points), stainless steel (light blue triangles), brass (green diamonds) and aluminium (red squares) plates. Lines conecting points serve only as eye guides. For reference the signal level recorded in the absence of the sample is shown with a solid black line. The plates are tilted by $20^{\circ}$ with respect to the primary rf field direction. The primary rf field coil is located above the centre of the plate.}\label{fig:Frequency_scan}
\end{figure}

Figure ~\ref{fig:Frequency_scan} shows the amplitude and phase of the rf signal as a function of the rf field frequency for ferrite (dark blue points), stainless steel (light blue triangles), brass (green diamonds) and aluminium (red squares) plates. For reference the signal recorded in the absence of an object is also shown (black solid line). A $20^{\circ}$ tilt between the normal to the plate surface and the primary rf field ensures that the component of the secondary field created by the eddy currents is visible.

Similarly to the angle scan, the set of points representing the amplitude and phase of the signal observed over the ferrite plate overlaps with that observed in the absence of a sample. As pointed out before, non-zero signal amplitude in the absence of the object results from residual misalignment between the primary rf field and sensor axes. The decrease of the signal amplitude with operating frequency is consitent with a similar dependence observed in a standard rf spectroscopy arrangement. It is useful to take the data recorded over ferrite and aluminium plates as the points of reference. Analysis of the frequency dependencies of the stainless steel and brass amplitudes and phases relative to ferrite and aluminium indicates the presence of three frequency regimes. The first, up to $\SI{4}{\kilo\hertz}$, represents frequencies where low induction efficiency results in low eddy current density. In this range the amplitude of the stainless steel overlaps with ferrite. For the frequencies in a second range, $\SI{4}{\kilo\hertz}$ - $\SI{15}{\kilo\hertz}$, a transition is observed in the stainless steel signal amplitude and phase from the level observed over ferrite to that recorded over the aluminium. In the third frequency range, above $\SI{15}{\kilo\hertz}$, all observed amplitude and phase values are close to their asymptotic levels. It is worth pointing out that the $\SI{29}{\kilo\hertz}$ operating frequency used to acquire the inductive images in  Fig. ~\ref{fig:Tilt_image} lies in this third frequency range, where the frequency dependence of the signals is negligible. It is worth comparing the frequency dependence of the phase changes of the signal generated by brass and stainless steel objects. The phase measured with brass, although smaller in value, mirrors the dependence observed for aluminium across the whole frequency range. These phase changes indicate that the inductive properties are dominated by electrical conductivity. With stainless steel the phase behaviour is similar to that of the ferrite at low frequencies, but approaches that of aluminium at higher frequencies.  This is consistent with an object that has significant  magnetic permeability and electrical conductivity.

\begin{figure*}[h!]
\includegraphics[width=\textwidth]{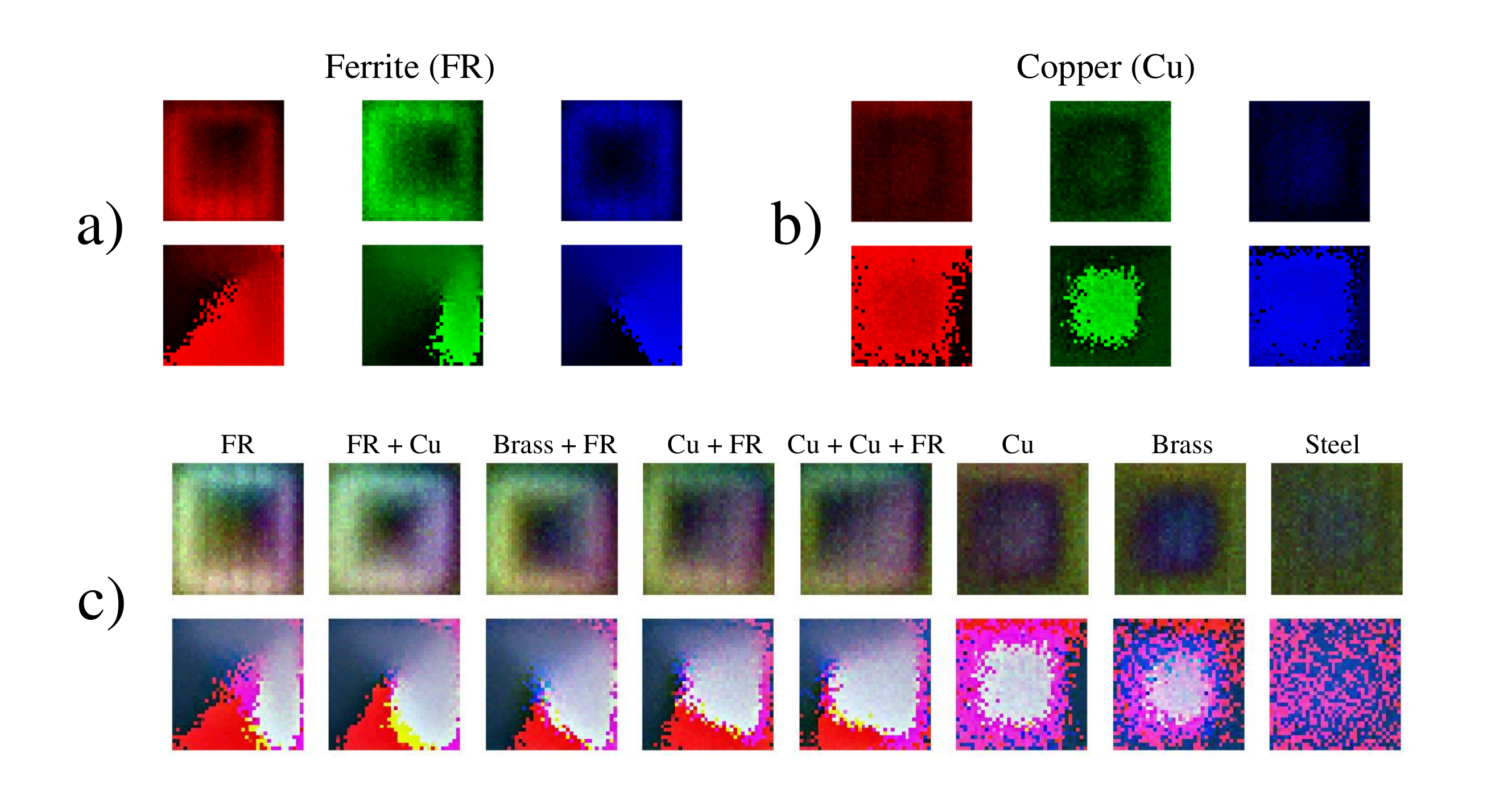}
\caption{The measured change of the amplitude (a) and phase (b) of the rf signal recorded over the ferrite and aluminium $35\times\SI{35}{\milli\meter\squared}$ plates at $\SI{3.5}{\kilo\hertz}$ (red), $\SI{7}{\kilo\hertz}$ (green), and $\SI{29}{\kilo\hertz}$ (blue). (c) Images integrated over three frequencies for various types of plates. }\label{fig:Image}
\end{figure*}

\section{Object mapping - spatial scan}

In this section we present two methods of the inductive image analysis. The first method uses the frequency dependence of the object response, whilst the second uses the integrated image amplitude.

We have previously shown that for conductive objects an optimum value within a 1–2 kHz frequency range can be identified, which maximises the amplitude and contrast of features (defect, edge signatures) observed in the inductive images \cite{Bevington2020a}. Similar behaviour was seen in magnetically permeable objects, but where the optimum values were shifted to a higher frequency range \cite{Bevington2021}. These observations suggest that monitoring the frequency dependence of the inductive image amplitude or contrast may indicate the object composition. The first analysis method follows the concept of colour perception by the human eye. White colour is a mixture of three basic (RGB) colours. Imbalance in these colour intensities will produce colour tones. 
In order to explore this capability, we have recorded images showing the amplitude and phase of the rf signal over either individual or stacks of $35\times\SI{35}{\milli\meter\squared}$ plates made of various materials. The images were recorded in a measurement configuration with the sensor axis parallel to the normal to the plate surface. In this configuration the non-zero signal is created solely by the edges of the object \cite{Bevington2020d}. For each object (i.e. plate or ensemble of plates) we recorded a set of images at $\SI{3.5}{\kilo\hertz}$, $\SI{7}{\kilo\hertz}$, and $\SI{29}{\kilo\hertz}$ and used them as the basis for an RGB representation. The images shown in Fig. ~\ref{fig:Image} (a)-(b) show scans over pure ferrite and copper plates recorded at $\SI{3.5}{\kilo\hertz}$ (red), $\SI{7}{\kilo\hertz}$ (green) and $\SI{29}{\kilo\hertz}$ (blue). Each of the frequency values used in this measurement represents one of the three frequency ranges identified in the previous section. The images within each set were normalised to the maximum amplitude value recorded within the set and summed up. 

Figure ~\ref{fig:Image} (c) shows the images integrated over three frequencies (RGB representation) for various other plates and sets of plates. A simple visual analysis of the images can be done by taking the ferrite and copper plates as the reference points. In this context, one can see the image of copper-ferrite plates set in Fig. ~\ref{fig:Image} (c) is a clear combination of the two references. Moreover, measuring two sheets of copper instead of one causes the conductivity fingerprint to be more pronounced, which is a manifestation of the layer's thickness. 
Because of the surface character of the effects in electrically conductive objects the amplitude of the image reflects the order of the materials in sets of plates.
Reversing the order of the ferrite and copper layers does not significantly modify the structure of the image but does lead to a colour change. When copper is on the top it screens the rf field and shifts the colour palette towards the copper plate.

\begin{figure*}[h!]
\includegraphics[width=\textwidth]{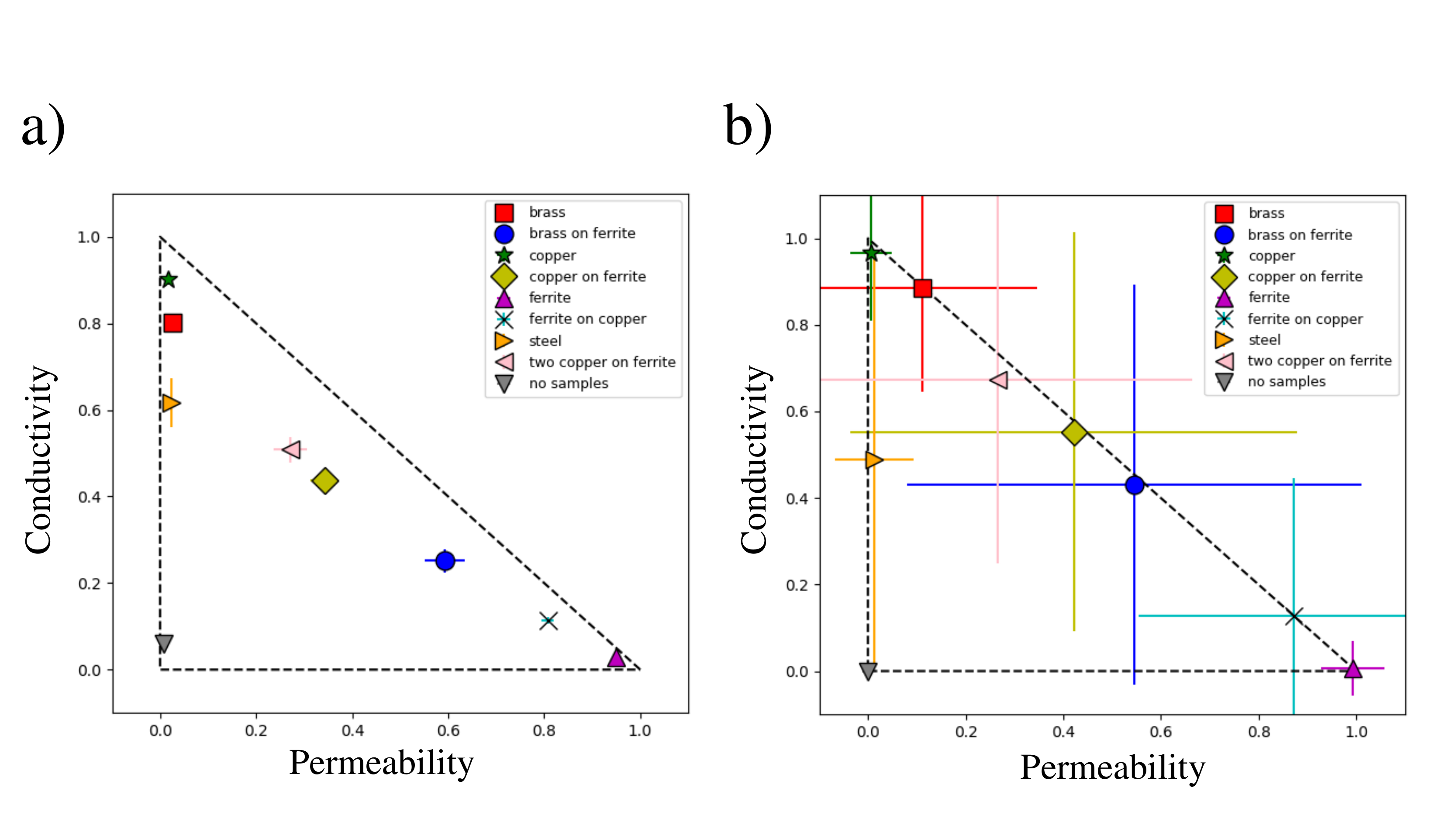}
\caption{Analysis of the object composition based on the signal amplitude integrated over the image area. The inductive images of the same plates as shown in Fig. ~\ref{fig:Image}, were recorded at $\SI{29}{\kilo\hertz}$. Each image (object) is represented by a point in 3D space defined by three orthogonal vectors that represent purely magnetically permeable (1,0,0-horizontal axis), electrically conductive (0,1,0- vertical axis) and background 'no-object detected' case (0,0,1- axis orthogonal to the plot plane). The position of the point shown in the plot is given by probabilities of the tested object having the reference properties (purely magnetically permeable, purely electrically conductive object, and no object detected). (a)/ (b) Distribution obtained with the custom metrics defined in the text/ standard the deep learning model. Error bars correspond to the standard deviation each set of samples.}\label{fig:Metric}
\end{figure*}

The second approach to inductive image analysis is based on the signal amplitude integrated over the image area. The method takes advantage of the opposite directions of the secondary fields created by the eddy currents and object magnetisation relative to an external reference such as a background field. Since the signal amplitude in the recorded image includes both the secondary and background fields, its total magnitude provides information on the relative orientation of these two components. Integration over an image area that includes elements like tilted surfaces, edges, etc. is in a sense analogous to the measurement of the angular dependence of the signal and can provide useful information for discrimination between object compositions. The method evaluates a measure of the probability that the object properties are the same as those of three reference standards: ferrite (an approximation of a purely magnetic object), copper (purely conductive object) and the absence of an object (plate presence). The integrated image amplitudes of these references define three vectors of an orthogonal basis for 3D space. Location in this space of the point representing a tested object is identified by three coordinates specified by metrics (proximity) with respect to the three references. This location is a measure of the probability of seeing the object, and of the object being electrically conductive or magnetically permeable.    
We introduce the metrics, which are a measure of proximity, d, of the given data point (tested object) to the specific reference standard x, as: $d_x=1/ \sum\limits_{i=1}^N {(R^i-R_{x}^{i})^2}$, where $R^i$ is the amplitude of a single pixel, $i$, in the tested object image, $R_{x}^{i}$ is the amplitude of the corresponding single pixel in the reference image, and N is the number of pixels in the image. The index $x$ refers to either the purely magnetically permeable (1,0,0), purely electrically conductive (0,1,0) and no plate case (0,0,1). The signal amplitudes in the reference image, $R_{x}^{i}$, are calculated as an average over 70 recorded images. Since the result, i.e. the data point location in 3D space represents probability, the sum of its coordinates is normalised, $d_x +d_y+d_z=1$. 

It is worth discussing the metrics structure in more detail, particularly the choice of the inverse dependence on amplitude difference, $(R^i-R_{x}^{i})^2$. The amplitude difference decreases as the object's properties become more similar to those of the reference. This leads to an increase in the value of the inverted factor, $1/(R^i-R_{x}^{i})^2$, which eventually becomes dominant over the other two factors, ie. the proximity to the other two references. Normalisation of the sum of the coordinates and projection of its position on the $xy$ (magnetic permeability vs electric conductivity) plane, places the measured point under the line $x+ y = 1$, in other words inside the triangle confined by $x= 0, y = 0$ and $x + y = 1$. The smaller the integrated amplitude difference is, the higher its inverted value and the closer its normalised value approaches 1. We have tested different types of metrics, in particular a linear dependence on integrated amplitude difference, as well as metrics including signal phase. While we have verified that all metrics provide similar qualitative results, i.e. spatial distribution of the tested points relative to reference standards, the metrics described here deliver the best differentiation between different materials. 

Figure ~\ref{fig:Metric} (a) shows the location of the points representing different materials (plates, sets of plates) in the 2D electrical conductivity - magnetic permeability plane. This subspace is chosen as it enables us to demonstrate the discrimination of objects based on composition. As a result of the normalisation condition the distance from the origin reflects the probability that an object is present, and hence is a demonstration of our ability to detect objects. Each point in the plot is an average over 15 images. The points are grouped near the line connecting purely magnetic objects (1,0) and purely electrically conductive objects (0,1). This indicates all the tested samples showed a significant degree of conductivity or permeability. It is worth pointing out that the method can distinguish between the set of plates with ferrite on top of copper (blue cross) and the same set in opposite order (yellow diamond). 
The images used in the measurements shown were recorded at $\SI{29}{\kilo\hertz}$, but equivalent data taken at $\SI{7}{\kilo\hertz}$ showed similar behaviour.  Data recorded at $\SI{3.5}{\kilo\hertz}$ were more scattered and led to poorer material discrimination, which is consistent with the observed weaker inductive signals at low frequencies. The similar distribution of the points in the data sets recorded at $\SI{7}{\kilo\hertz}$ and $\SI{29}{\kilo\hertz}$ indicates that even an image recorded at one frequency may contain enough information for the discrimination of object composition.

The above procedure relies on the fact that all the objects have the same shape and position within the image. More flexible alternatives could be use in the form of machine learning. To test this we implemented a convolutional neural network constructed from a combination of standard layers applied in computer vision tasks (convolutional layer, pooling layer, dense layers and activation layer). The advantage of the algorithm is its ability to make a decision based on the fragments of the whole sample. The model was trained on 8x8 pixel fragments cut from 17x17 pixel images. This allowed us to increase number of 230 images available for training by factor of 100. Similar fragments of scans were then used to test trained model predictions for unknown sample types. Averaged results for all samples are presented in Fig. ~\ref{fig:Metric} (b). Individual points are strongly scattered which resulted in bigger uncertainties than in Fig. ~\ref{fig:Metric} (a) but the distributions of points are similar. 
Relatively large uncertainties in Fig. ~\ref{fig:Metric} (b) are caused by the small number of object categories used for the training (free space, ferrite, copper), which was equal to the number of the properties (free space, ferromagnetic, conductor). Increase in the number of object categories (predefined standards) used in the training process would significantly increase model accuracy even if these categories did not cover all types of objects expected in tests. In other words, the ability of the algorithm to identify object properties with small uncertainty will be enhanced by introducing more standards with similar, but not necessarily the same, qualities/characteristics.

The ability to discriminate between ferrite plates and a mixture of copper and ferrite plates (Fig. ~\ref{fig:Metric}) shows that the combination of the measurement geometry and the difference in angular responses generated by eddy currents and magnetisation (Fig.~\ref{fig:Secondary_field}) allows us to see objects behind barriers or within electrically conductive enclosures.

An important concern is the practicality of implementing the presented methods in object screening. The results shown in Fig. ~\ref{fig:Metric} were recorded with objects that have the same geometry and dimensions, which is a highly idealized case. 
One possible approach would be the use of a geometry non-specific procedure that combines large-scale inductive imaging of an object followed by the identification of appropriate features for composition analysis. The challenge of comparing results from objects with different geometries can also be addressed by more powerful tools such as supervised and unsupervised machine learning methods, which have proven to be very successful tools in solving similar problems \cite{Wu2020}. The implementation of machine learning used to generate the results in  Fig. ~\ref{fig:Metric} (b) was successful despite using only the amplitude of the measured signals at a single frequency. Enhanced performance would be expected from an implementation incorporating a combination of frequency, spatial and angle data.

The main aim of this paper is the demonstration that a combination of the three degrees of freedom (spatial, angular and frequency) applied in the inductive measurements can provide sufficient information to deduce object composition. We anticipate that similar information combined with advanced machine learning techniques will provide an even more versatile and effective tool, in which an optimised measurement sequence for a specific implementation is determined by the machine controlling the process. In this scenario the actual test would consist of a series of moves using all the degrees of freedom in a sequence that is autonomously decided and continuously updated by a pretrained machine learning model that, at the end of the measurement procedure, would provide some specific information about the interrogated object based on the collected data.

\section{Conclusions}
In conclusion, we have demonstrated a series of MIT measurements that can assist in the identification of an object's composition. We showed that the angular, frequency and spatial dependencies of the rf signal recorded over the objects can discriminate between objects made of materials with different magnetic permeability and electrical conductivity. The concept relies on the different penetration depth in purely electrically conductive and magnetically permeable materials. The skin depth that reflects this penetration depth is a function of the operating frequency, electrical conductivity, and magnetic permeability. Observed signal dependences confirm that in electrically conductive materials the secondary field is created at the surface. This could explain why permittivity does not play a significant role in our measurements. The discrimination is made possible through the use of a sensor with an insensitive axis. This eliminates the contribution of the primary rf field to the signal and gives the sensor a different sensitivity to the eddy current and magnetisation components of an induced response. We have discussed the influence of object shape on the angular dependence of the rf signal. Whilst frequency scans can be performed at a single location over an object, the measurement of angular dependence requires a physical change of the measurement configuration. This could be performed in various ways. In the specific case of goods screening, the objects are often transferred with a conveyer belt. Location of a sensor at a bend in such a system would allow the measurement of objects at similar distances but at different orientations relative to the sensor axis. Finally, we have demonstrated that even very simple methods for acquiring and analysing inductive images can successfully discriminate between different materials.

We acknowledge the support of the UK government department for Business, Energy and Industrial Strategy through the UK national quantum technologies programme.

The data that support the findings of this study are available from the corresponding author upon reasonable request.

\end{document}